\documentclass[10pt]{iopart}

\usepackage{color,graphicx}
\usepackage{subfig}
\usepackage{braket}

\begin{document}

\title[]{Exploring the Kibble-Zurek mechanism with homogeneous Bose gases}

\author{J Beugnon$^1$ \& N Navon$^2$}

\address{$^1$ Laboratoire Kastler Brossel, Coll\`ege de France, CNRS, ENS-PSL Research University, UPMC-Sorbonne Universit\'es,
11 place Marcelin-Berthelot, 75005 Paris, France}
\address{$^2$ Cavendish Laboratory, University of Cambridge, J. J. Thomson Avenue, Cambridge CB3 0HE, United Kingdom}

\ead{beugnon@lkb.ens.fr, nn270@cam.ac.uk}

\vspace{10pt}
\begin{indented}
\item[]\today
\end{indented}

\begin{abstract}
Out-of-equilibrium phenomena is a subject of considerable interest in many fields of physics. Ultracold quantum gases, which are extremely clean, well-isolated and highly controllable systems, offer ideal platforms to investigate this topic. The recent progress in tailoring trapping potentials now allows the experimental production of homogeneous samples in custom geometries, which is a key advance for studies of the emergence of coherence in interacting quantum systems. Here we review recent experiments in which temperature quenches have been performed across the Bose-Einstein condensation (BEC) phase transition in an annular geometry and in homogeneous 3D and quasi-2D gases. 
Combined, these experiments give a comprehensive picture of the Kibble-Zurek (KZ) scenario through complementary measurements of correlation functions and topological defects density. They also allow the measurement of KZ scaling laws, the direct confirmation of the ``freeze-out" hypothesis that underlies the KZ theory, and the extraction of critical exponents of the Bose-Einstein condensation transition.

\end{abstract}

\ioptwocol

\section{Introduction}\label{sec1}
Out-of-equilibrium physics of quantum many-body systems is the focus of a wide range of both experimental and theoretical studies \cite{Eisert15}. A cornerstone of out-of-equilibrium physics is the Kibble-Zurek theory, an elegant framework that describes the emergence of an ordered phase after a quench through a second-order phase transition, and consequences of the critical dynamics of spontaneous symmetry breaking, such as the formation of topological defects. The inevitability of defect production when crossing a phase transition was first noted in the context of cosmology \cite{Kibble76}, where relativistic causality yielded a lower bound on defect density. The role of critical slowing down in second-order phase transitions and the connection with universality classes was introduced in \cite{Zurek85}, that also suggested test of cosmological ideas in condensed-matter systems.
Since these seminal proposals, the Kibble-Zurek mechanism (KZM) has been studied in many different media such as liquid helium \cite{Ruutu96,Bauerle96}, superconducting loops \cite{Monaco09}, ferroelectric materials \cite{Lin14}, liquid crystals \cite{Chuang91}, colloidal monolayers \cite{Deutschlaender15} or trapped ions \cite{Pyka13,Ulm13,delCampo10}. First developed for classical phase transitions, the KZM has also been extended to the case of quantum phase transitions \cite{Zurek05,Dziarmaga05,Dziarmaga10,Chen11,Baumann11,Braun15,Clark16}. Several reviews give an extensive account of past experimental developments \cite{Kibble07,Zurek96,delCampo14}.

Ultracold gases are fascinating systems to investigate the KZM. They are well-controlled and well-isolated from their environment, and their equilibrium states have been studied in detail. The atomic physicist toolbox allows one to create out-of-equilibrium situations by, for instance, varying in time the trapping potential or the strength of inter-particle interactions. The first investigations of Kibble-Zurek physics with ultracold atoms were done with harmonically trapped gases, via the production of topological defects \cite{Weiler08,Lamporesi13,Donadello16}. The statistical study of these defects in \cite{Lamporesi13,Donadello16} enabled the extraction of critical exponents of the phase transition but it required the theoretical development of a modified KZM, which takes into account the non-uniform gas density \cite{Zurek09,Damski10}. Ultracold atoms can also be used to study the KZM in systems in multi-component Bose-Einstein condensates \cite{Sadler06,Damski08,Damski09,Sabbatini11,Swislocki13,De14,Anquez16,Liu16}. 

Recent experimental advances now enable the production of essentially uniform quantum gases. In this review, we focus on a set of recent experiments investigating the KZM with uniform Bose gases \cite{Corman14,Navon15,Chomaz15}. These experiments offer a new perspective thanks to their specific geometries. Using tailored optical potentials, they implement custom-shaped ``flat-bottom" atom traps. We review here experiments realized with a 3D box potential \cite{Navon15}, an annulus \cite{Corman14} and a quasi-2D uniform potential \cite{Chomaz15}. In all cases quench cooling is applied to the cloud and either correlation functions are measured or topological defects are probed. The experimental findings are in agreement with the Kibble-Zurek scenario and allow the extraction of the critical exponents of the Bose-Einstein condensation (BEC) phase transition.

The review is organized as follows. Section 2 briefly introduces the basics of the KZM: the critical slowing down phenomenon, which is a central assumption of the KZM, and the different regimes in the time evolution of the system. Section 3 is devoted to the experimental realization of uniform quantum gases. Section 4 focuses on the tools used to probe the system at different stages of its evolution. 
Section 5 presents the determination of critical exponents and in the last section we discuss future perspectives.

%%%%%%%%%%%%%%%%%%%%%%%%%%%%%%%%%%%%%%%%%%%%%%%%%%
%%%%%%%%%%%%%%%%%%%%%%%%%%%%%%%%%%%%%%%%%%%%%%%%%%
%%%%%%%%%%%%%%%%%%%%%%%%%%%%%%%%%%%%%%%%%%%%%%%%%%
%%%%%%%%%%%%%%%%%%%%%%%%%%%%%%%%%%%%%%%%%%%%%%%%%%
%%%%%%%%%%%%%%%%%%%%%%%%%%%%%%%%%%%%%%%%%%%%%%%%%%
%%%%%%%%%%%%%%%%%%%%%%%%%%%%%%%%%%%%%%%%%%%%%%%%%%

\section{The Kibble-Zurek Mechanism}\label{sec2}

Consider a system described, in equilibrium, by a correlation length $\xi$. In the vicinity of the critical point of a second-order phase transition this quantity diverges as a power law. For a classical phase transition, the divergence of $\xi$ is characterized by the static critical exponent $\nu$
\begin{eqnarray}
\xi \propto |T-T_c|^{-\nu},
\label{eq1}
\end{eqnarray}
where $T$ is the temperature, and $T_c$ is the phase transition's critical temperature. 
The relaxation time $\tau$ needed to establish these diverging correlations, also diverges and is characterized by the dynamical critical exponent $z$:
\begin{eqnarray}
\tau \propto \xi^z.
\label{eq1b}
\end{eqnarray}
It is thus increasingly harder to establish the equilibrium state as one approaches the phase transition point, a phenomenon called critical slowing down. 

\begin{figure}[hbt!!]
\centering
\includegraphics[width=\columnwidth]{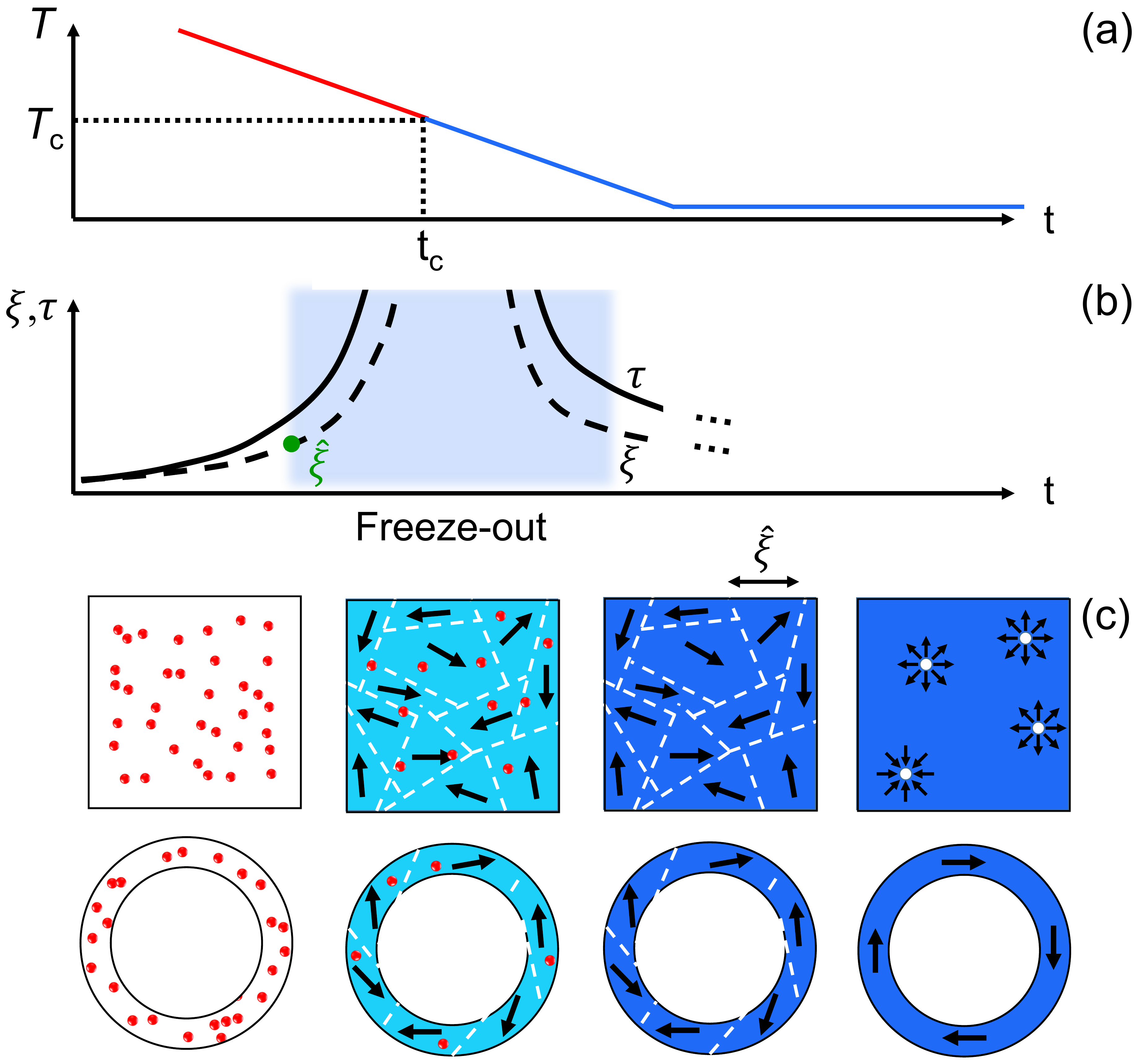}
\caption{The Kibble-Zurek mechanism. (a) Linear cooling ramp $T(t)$, where the critical temperature $T_c$ is crossed at the time $t=t_c$. (b) Schematic of the evolution of the equilibrium correlation length $\xi(t)$ (dashed line) and relaxation time $\tau(t)$ (solid line) during the temperature ramp. In the region around the critical point where $\tau(t)>|t-t_c|$ (shaded area) the system is `frozen' and the correlation length is fixed to a value $\hat{\xi}$ (green point). (c) Cartoons describing the state of the system at different times along the evolution, both in a bulk and an annulus geometry (the orientation of the arrows - unit vectors - encodes the local phase of the order parameter).
Particles in the disordered phase are shown as red dots, a small (large) ordered-phase fraction is shown in light (dark) blue. In the freeze-out region, domains of ordered phase with independent phases are formed. After the freeze-out time the system coarsens, leading to the stochastic formation of topological defects such as point-like vortices, or supercurrents.}
\label{figKZ}
\end{figure}

The classical KZM models the time evolution of a system during the cooling process (shown in figure \ref{figKZ}a) into a sequence of three steps (figure \ref{figKZ}b). The first part is an adiabatic evolution in which the relaxation time is smaller than the characteristic evolution time of the cooling. The range of correlations then closely follows the instantaneous equilibrium value $\xi(t)$ during the cooling. When the relaxation time becomes larger than the remaining time before reaching the transition, the system enters the so-called ``freeze-out" period (blue shaded area in figure \ref{figKZ}b), and the correlation length $\xi$ is assumed to remain fixed at a value $\hat{\xi}$ (green point in figure \ref{figKZ}b). The system thus crosses the phase transition without $\xi$ ever having spanned the whole system. This results in the formation of broken-symmetry phase domains that have random independent values of the order parameter and typical size $\hat{\xi}$ (see figure \ref{figKZ}c). In the third and last period a subsequent evolution (e.g. coarsening) takes place.

Quantitatively, we assume a ramp of temperature $T(t)$ (with $T(0)>T_c$) that is linear around the critical point 
\begin{eqnarray}
T(t)/T_c=1+(t_c-t)/\tau_Q,
\label{eq2}
\end{eqnarray}
where $\tau_Q$ is the characteristic quench time, and the phase transition is reached at the time $t=t_c$. From these assumptions one can easily deduce a power-law scaling of the typical domain size with the quench time \cite{Zurek85},
\begin{equation}
\hat{\xi} \propto \tau_Q^{-\frac{\nu}{1+\nu z}}.
\label{eq3}
\end{equation}
The power-law exponent $\nu/(1+\nu z)$ depends only on the values of the critical exponents describing the equilibrium state.

In addition to the formation of ordered-phase domains, the KZM also predicts the nucleation of topological defects, resulting from the merging of domains with different values of the order parameter after the freeze-out period (see figure \ref{figKZ}c). Such defects can be long-lived and constitute a robust signature of the quench-cooling dynamics. 
The average number of defects $\langle N_d \rangle$ is also predicted to follow a power-law scaling with the quench time $\tau_Q$,
\begin{eqnarray}
\langle N_d \rangle \propto \tau_Q^{-\frac{\alpha\nu}{1+\nu z}},
\label{eq4}
\end{eqnarray}
where $\alpha$ is a coefficient that takes into account the dimensionality of the system and that of the defects \cite{delCampo14}. While the Kibble-Zurek description of the quench dynamics consisting of distinct steps is obviously a simplified picture of the actual critical dynamics, its predictions (\ref{eq3}) and (\ref{eq4}) can be tested experimentally, as well as its underlying assumptions. A complementary point of view suggests that, even though a sudden ``freezing" of correlations does not occur in practice, the velocity at which correlations spread is finite near the transition point and consequently, the domains are bound to form within a sonic horizon of size $\propto \hat{\xi}$, in agreement with the KZM \cite{Zurek85,Francuz16}.

We now consider the specific case of the Bose-Einstein phase transition for weakly interacting particles. It is a second-order phase transition between a thermal state and a Bose-condensed, superfluid state. The order parameter for the Bose-Einstein transition is the condensate wavefunction and the system exhibits long-range order below $T_c$. The symmetry broken by the Bose-condensed phase is a U(1) symmetry, which corresponds to an arbitrary choice of the condensate wavefunction's phase.

%%%%%%%%%%%%%%%%%%%%%%%%%%%%%%%%%%%%%%%%%%%%%%%%%%
%%%%%%%%%%%%%%%%%%%%%%%%%%%%%%%%%%%%%%%%%%%%%%%%%%
%%%%%%%%%%%%%%%%%%%%%%%%%%%%%%%%%%%%%%%%%%%%%%%%%%
%%%%%%%%%%%%%%%%%%%%%%%%%%%%%%%%%%%%%%%%%%%%%%%%%%
%%%%%%%%%%%%%%%%%%%%%%%%%%%%%%%%%%%%%%%%%%%%%%%%%%
%%%%%%%%%%%%%%%%%%%%%%%%%%%%%%%%%%%%%%%%%%%%%%%%%%

\section{Realization of homogeneous ultracold gases}\label{sec3}

Ultracold gases can be trapped in optical dipole traps  formed by far-detuned laser beams \cite{Grimm00}. 
In the works described here, the laser light is blue-detuned with respect to the atomic resonance of interest and atoms are thus repelled from regions of high laser intensity. The initial energy depth of these traps is in the range of $k_B \times$(1--10)\,$\mu$K, much larger than the atoms average energy around the critical point. The profile of the laser beams can be shaped using (for instance) holographic phase plates \cite{Rath10}, masks \cite{Krinner15}, high speed deflectors \cite{Zimmermann11} and programmable spatial light modulators (SLM) \cite{Bergamini04}. Atoms are confined on typical sizes of a few tens of microns. 
Quench cooling is performed by lowering the power of the laser beams forming the trapping potential. The reduced trap depth leads to forced evaporative cooling of the cloud.

\begin{figure}%
    \centering
    \subfloat{{\includegraphics[width=\columnwidth]{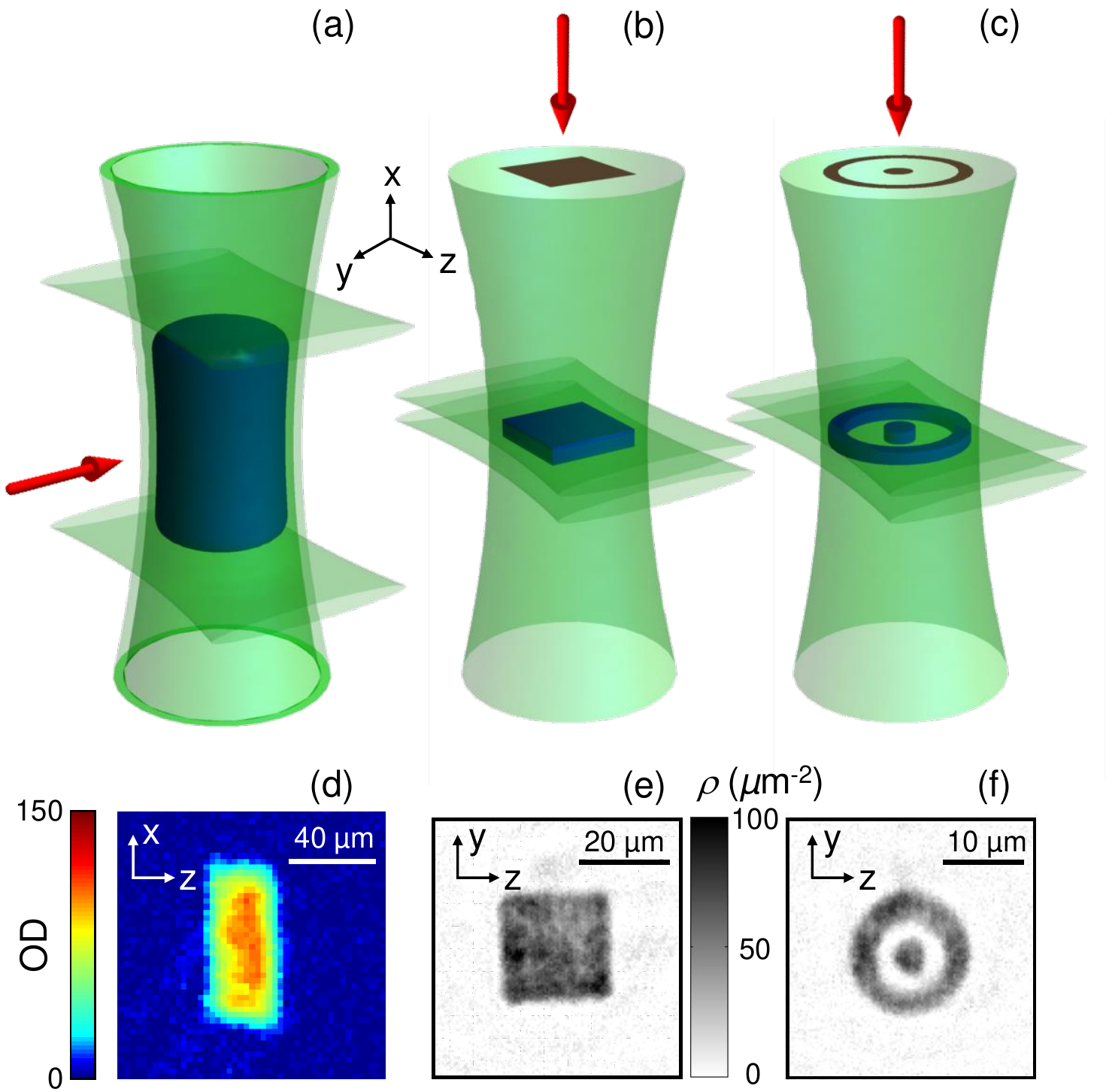}}}
		
\caption{Uniform atomic traps. (a-c) Cartoons of the trapping potentials. (a) Cylinder-shaped 3D box \cite{Navon15}, (b) square 2D box \cite{Chomaz15}, (c) ``target" trap \cite{Corman14}. The trapping laser light is represented in green,  the atomic gas, in blue, and in (b)-(c) the masks are shown in brown. The red arrows indicate the imaging axis. (d-f) Examples of \textsl{in situ} absorption images in (d) a 3D cylindrical box of radius of $16$ $\mu$m and length of 65 $\mu$m \cite{Gaunt13}, (e) a square trap of size 24\,$\mu$m, and (f) a ``target" trap  with a central disk of radius 4.5\,$\mu$m and an outer ring of radii 9\,$\mu$m and 15\,$\mu$m. From \cite{Corman14,Gaunt13}, \copyright American Physical Society.}
\label{fig1}
\end{figure}

In figure \ref{fig1} we present the three geometries discussed in this review: a 3D uniform gas, a quasi-2D uniform gas and a ``target" configuration where atoms are confined in a quasi-2D ring and a central disk. The 3D gas is realized using a SLM to shape an incoming gaussian beam into a hollow ring and two thin end caps which respectively form the tube and the walls of a cylinder-shaped trap. In \cite{Navon15}, the 3D box potential has a radius $R\approx 17$ $\mu$m, and a length $L\approx 26$ $\mu$m, and the typical critical temperature for BEC is $170\,$nK. In the two other configurations \cite{Corman14,Chomaz15}, the cloud is strongly confined along one direction ($x$ axis in figure \ref{fig1}), in the node of the Hermite-Gauss laser beam \cite{Rath10}. The oscillation frequency along this direction is $365\,$Hz. The in-plane trapping is realized by imaging a dark mask with custom shape onto the atomic cloud \cite{Chomaz15}. 

%%%%%%%%%%%%%%%%%%%%%%%%%%%%%%%%%%%%%%%%%%%%%%%%%%
%%%%%%%%%%%%%%%%%%%%%%%%%%%%%%%%%%%%%%%%%%%%%%%%%%
%%%%%%%%%%%%%%%%%%%%%%%%%%%%%%%%%%%%%%%%%%%%%%%%%%
%%%%%%%%%%%%%%%%%%%%%%%%%%%%%%%%%%%%%%%%%%%%%%%%%%
%%%%%%%%%%%%%%%%%%%%%%%%%%%%%%%%%%%%%%%%%%%%%%%%%%
%%%%%%%%%%%%%%%%%%%%%%%%%%%%%%%%%%%%%%%%%%%%%%%%%%

\section{Experimental probes of the KZM }

In this section, we present experimental tools that enable the probing of several consequences of the KZM, at different stages of a quench. When the system is probed within the time interval in which the correlations are frozen by the critical slowing down, the central prediction (\ref{eq3}) of the KZM should apply, namely that the order parameter exhibits correlations on a scale given by the value of the (frozen) correlation length $\hat{\xi}$. In subsection \ref{subsec41} we show that the range of the correlations can be probed via homodyne interferometry. Importantly, the freeze-out hypothesis can be tested with the same tool. 

After the freeze-out period, the system will resume its evolution, the extent of the correlations will change, and the ordered-phase domains will follow coarse-graining dynamics. Interestingly, this phenomenon leads to the formation of defects whose density is still comprehensible within the KZM framework (\ref{eq4}). 
In subsection \ref{subsec42}, we describe their observation via direct absorption imaging in the case of a 2D geometry, or via the detection of the phase winding in an annular geometry.

\subsection{First-order correlation function and the freeze-out hypothesis}\label{subsec41}

\begin{figure*}[hbt!!!]
\centering
\includegraphics[width=2\columnwidth]{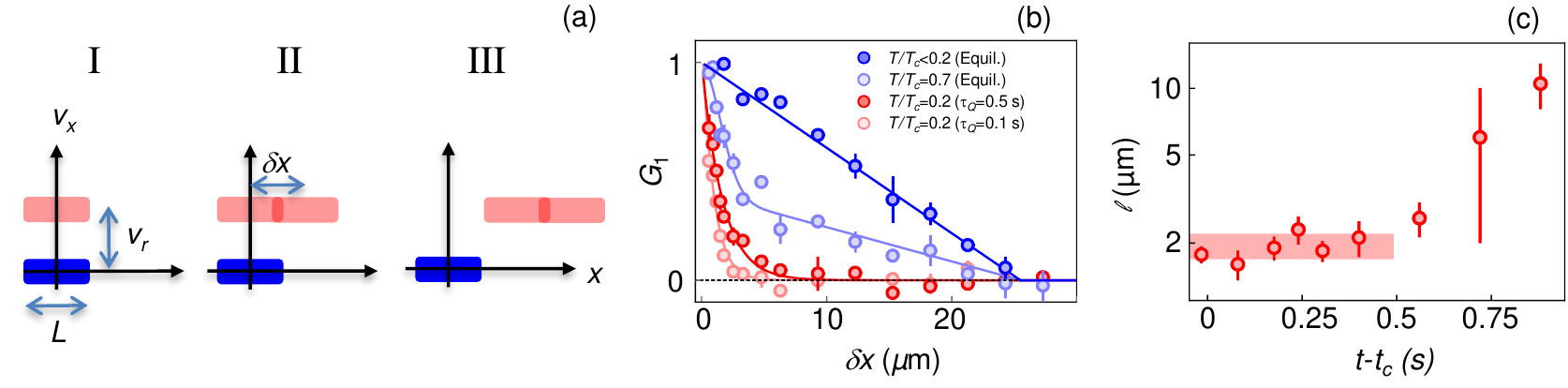}
\caption{Measurement of the first-order correlation function. (a) Sketch of the interferometric scheme to probe $g_1$. I: a superposition of the cloud at rest (blue area) and moving at $v_r$ (red area) is created. II: the moving part is displaced by $\delta x$ and allowed to interfere with the original copy. III:  the resulting fraction moving at $v_r$ is separated from the one at rest for measurement. (b) Examples of correlation functions measured in a box of size $L\approx 26$ $\mu$m: a nearly pure condensate (dark blue), a partially condensed gas (light blue), and quenched samples for two different quench times (red points). (c) The freeze-out hypothesis: $\ell$ versus observation time after the quench $t-t_c$ for a total cooling time $t_Q=3.2$ s, and $t_c=2.3$ s. The vertical (resp. horizontal) extent of the shaded area represents the average $\ell$ (resp. freeze-out time) for this quench. (for the nonlinear cooling trajectories used in \cite{Navon15}, $\tau_Q\approx 0.4\;t_Q$). From \cite{Navon15}, \copyright Science AAAS.}
\label{figG1}
\end{figure*}

The range of correlations can be extracted from the first-order correlation function, $g_1({\bf x},{\bf x'})\propto\langle \hat{\Psi}^\dag({\bf x})\hat{\Psi}({\bf x'})\rangle$ (where $\hat{\Psi}$ is the field operator), which for a translationally-invariant system is a function of $|{\bf x}-{\bf x'}|$ only. 
A simple method to measure $g_1$ in cold atomic systems was developed in \cite{Hagley99,Navon15} and is depicted in figure \ref{figG1}a. 
Using a short pulse of Bragg-diffraction beams, one creates a coherent superposition of the system with a center of mass motion at rest, and moving at a recoil velocity $v_r$ along the $x$ axis of the tube (see figure \ref{fig1}a). After a wait time $\delta t$, the moving component of the superposition is displaced by $\delta x=v_r \delta t$ (see figure \ref{figG1}a). Using a recombining pulse, the two components interfere and the total fraction moving at $v_r$ depends on the relative phase of the two components in the overlapping region. The total fraction diffracted after the second pulse for a system trapped in a box of length $L$ is \cite{Navon15}: 
\begin{equation}
\frac{N_r}{N}(\delta x)\propto 1+\left(1-\frac{\delta x}{L}\right)g_1(\delta x).
\end{equation}

In figure \ref{figG1}b, examples of experimental correlation functions $G_1(\delta x)\equiv (1-\delta x/L)g_1(\delta x)$ are shown. For a quasi-pure condensate in equilibrium, $g_1\approx 1$, $G_1\approx 1-\delta x/L$, and the correlations extend up to the system size $L$ ($\approx 26$ $\mu$m in figure \ref{figG1}b). For a partially condensed cloud (light blue), the presence of short-range thermal correlations results in a fast decay of $G_1$ for small $\delta x$, but the condensed part remains phase coherent over the entire box. 

When the Bose gas is (rapidly) quenched, the system instead exhibits only short-range correlations (red points in figure \ref{figG1}b), despite the low final temperatures $T/T_c\approx 0.2$ \cite{Navon15}. The extent of the correlations is larger for slower quenches, which is in qualitative agreement with the KZM. The data is well fitted by $\propto \exp(-\delta x/\ell)$, where the parameter $\ell$ characterizes the range of the correlations. 

The Kibble-Zurek scaling law (\ref{eq3}) is derived by neglecting post-quench dynamics. This naturally raises the question as to when the correlation function has to be measured in order to observe (\ref{eq3}). This can be addressed by varying the time when $\ell$ is measured along the quench trajectory. An example of such measurement is shown in figure \ref{figG1}c. One observes that for some time after the critical point has been crossed, $\ell$ is approximately insensitive to the measurement time $t-t_c$, thus providing direct support for the freeze-out hypothesis. If however $t-t_c$ is too large, $\ell$ rises and quickly reaches the system size.

%%%%%%%%%%%%%%%%%%%%%%%%%%%%%%%%%%%%%%%%%%%%%%%%%%
%%%%%%%%%%%%%%%%%%%%%%%%%%%%%%%%%%%%%%%%%%%%%%%%%%
%%%%%%%%%%%%%%%%%%%%%%%%%%%%%%%%%%%%%%%%%%%%%%%%%%
%%%%%%%%%%%%%%%%%%%%%%%%%%%%%%%%%%%%%%%%%%%%%%%%%%
%%%%%%%%%%%%%%%%%%%%%%%%%%%%%%%%%%%%%%%%%%%%%%%%%%
%%%%%%%%%%%%%%%%%%%%%%%%%%%%%%%%%%%%%%%%%%%%%%%%%%

\subsection{Formation, detection and lifetime of defects}\label{subsec42}

The formation of defects in Bose-Einstein condensates during post-quench dynamics results from the merging of domains with different order-parameter phases. For instance, in a bulk 2D system (figure \ref{figKZ}c), a state with a phase winding of $2\pi$ times an integer around given points is likely to appear for suitable arrangements of the relative phase of neighboring domains, corresponding to the nucleation of quantized vortices. In an annular geometry, the phase winding along the annulus corresponds to the formation of quantized superfluid currents \cite{Zurek85}. 

After the initial work of Zurek \cite{Zurek85}, it was proposed in \cite{Anglin99} and investigated numerically in \cite{Antunes99} that rapid  Bose-Einstein condensation can create vortices (or superfluid currents) in a trapped atomic Bose-Einstein condensate. On the experimental side, stochastic formation of vortices was observed after merging several independent Bose-Einstein condensates  \cite{Scherer07} or when quench cooling the gas \cite{Weiler08,Lamporesi13,Donadello16}.

\begin{figure}[hbt!!]
\centering
\includegraphics[width=8.2cm]{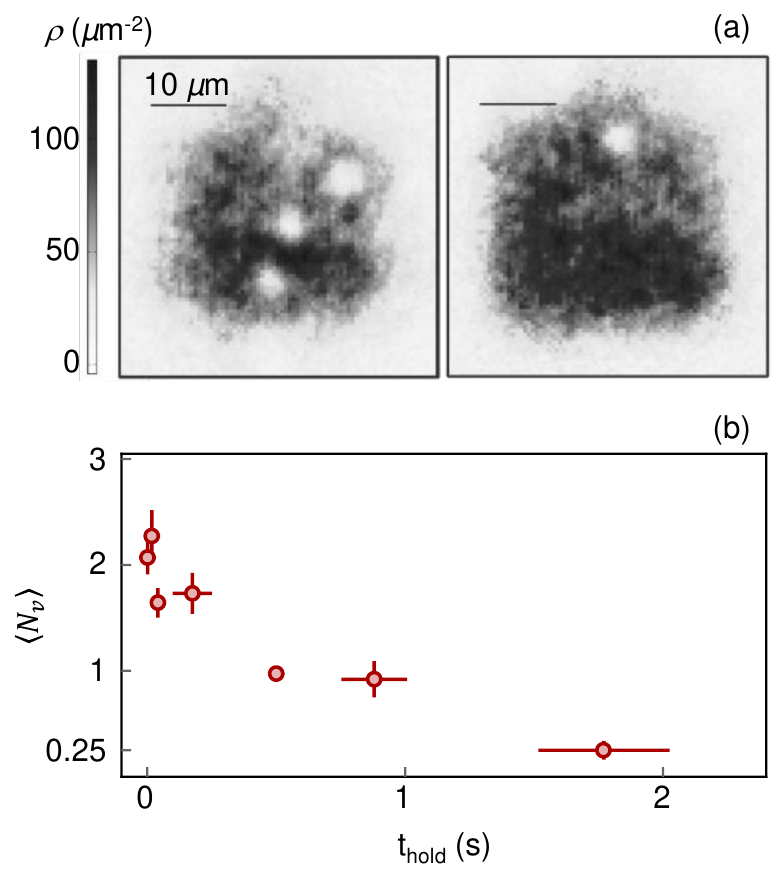}
\caption{Detection and lifetime of vortices generated after a quench. (a) Absorption images after a short time-of-flight expansion from the square-box trap. High-contrast holes in the density profiles correspond to bulk vortices in the cloud. The position of the vortices is stochastic and so is their number (respectively three and one in the two examples). (b) Mean vortex number $\langle N_v \rangle$ versus hold time $t_{\rm{hold}}$ after a quench of $t_Q=50$ ms. From \cite{Chomaz15}, \copyright Nature Publishing Group.}
\label{figvortex}
\end{figure}

In a superfluid, the size of a vortex core is given by the healing length which, for ultracold gases, is on the order of several hundreds of nanometers. It is thus not easy to detect them directly with \textsl{in situ} absorption imaging because of limited optical resolution. A short time-of-flight expansion (typically a few milliseconds), corresponding to the free evolution of the cloud without the trap, allows the vortex cores to expand and to become easily detectable with standard imaging techniques, as can be seen in figure \ref{figvortex}a. In \cite{Chomaz15} the stochastic character of the creation of vortices is confirmed by observing that the vortices position is random and uniformly distributed in the cloud.

\begin{figure}[hbt!!]
\centering
\includegraphics[width=\columnwidth]{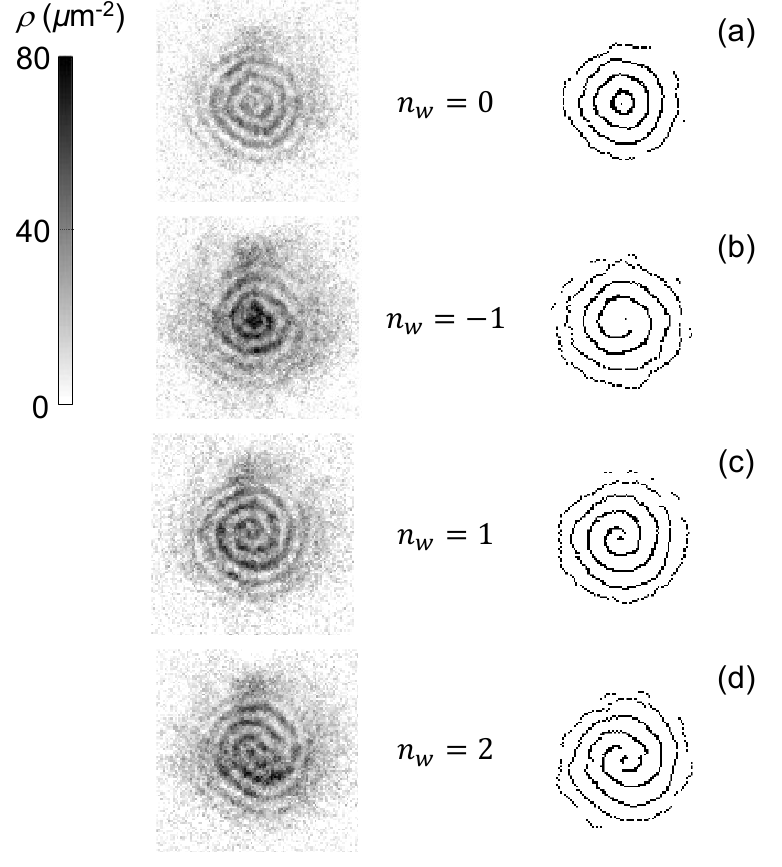}
\caption{Absorption images (left) and reconstructed fringe pattern (right) after a short in-plane time-of-flight from the ``target" trap. Atoms in the central disk and the outer ring interfere, resulting in either (a) concentric fringes or (b-d) spiral patterns corresponding to different winding numbers: -1 (b), +1 (c), and +2 (d). From \cite{Corman14}, \copyright American Physical Society.}
\label{figspiral}
\end{figure}

Superfluid currents along an annulus can also be detected in time-of-flight \cite{Moulder12,Murray13} but in the experiments described here with the ``target" configuration, an interference technique is used to detect simultaneously the amplitude and the direction of the superfluid current \cite{Corman14,Eckel14}. The basic idea is to let atomic clouds, initially confined in the annulus and in the inner disk, expand and interfere with each other. The inner disk is small enough so that vortices are unlikely to form there during quench cooling. This disk thus acts as a phase reference for the annulus. When no circulation is present in the annulus, concentric fringes are observed (see figure \ref{figspiral}a). When circulation is present in the annulus, spiral patterns are observed and the direction and the amplitude of the quantized superfluid current can be directly deduced from these pictures (see figure \ref{figspiral}b-d). By virtue of the symmetry of the annulus, the stochastically generated superfluid current has equal probability of having clockwise or counterclockwise direction. 
Defining the winding number $n_w$ as the number of quanta of circulation (which can  be positive, negative or null, as in figure \ref{figspiral}), an average winding number $\langle n_w \rangle=0.00(2)$ is measured for all the data in \cite{Corman14}. 

A motivation for studying topological defects stems from their topological character, that usually ensures a long lifetime. As they are easy to detect in quantum-gas experiments, they constitute a good probe of the critical dynamics during the quench cooling. However, there are limitations to this picture. Generally, the measured number of vortices is always smaller than expected from the number of initial phase domains \cite{Das12,Laguna97,Yates98}. This effect can result from merging dynamics \cite{Scherer07,Carretero-Gonzalez08}, the possibility for vortices with opposite circulation to annihilate, or the fact that single vortices can disappear when they reach the edge of the trap.  
As an example we show in figure \ref{figvortex}b the time evolution of the average number of vortices $\langle N_v \rangle$ in the square configuration and for a cooling time of 50\,ms. In this case, a lifetime of about 1\,s is measured. This lifetime is not large compared to some quench times, and to some hold times used before measurement (typically 500\,ms, to let other excitations dissipate and allow a robust determination of the number of vortices \cite{Biroli10}). On the other hand, the lifetime of the superfluid current in the ``target" configuration was measured to be around 7\,s (not shown here). These  latter defects are thus more robust and constitute a more reliable probe of the KZM. These effects can play an important role in the determination of the critical exponents based on (\ref{eq3}) and (\ref{eq4}), as discussed in the next section.

%%%%%%%%%%%%%%%%%%%%%%%%%%%%%%%%%%%%%%%%%%%%%%%%%%
%%%%%%%%%%%%%%%%%%%%%%%%%%%%%%%%%%%%%%%%%%%%%%%%%%
%%%%%%%%%%%%%%%%%%%%%%%%%%%%%%%%%%%%%%%%%%%%%%%%%%
%%%%%%%%%%%%%%%%%%%%%%%%%%%%%%%%%%%%%%%%%%%%%%%%%%
%%%%%%%%%%%%%%%%%%%%%%%%%%%%%%%%%%%%%%%%%%%%%%%%%%
%%%%%%%%%%%%%%%%%%%%%%%%%%%%%%%%%%%%%%%%%%%%%%%%%%

\section{Determination of critical exponents}\label{sec7}

An interesting feature of the KZM is the possibility to extract \emph{static} critical exponents from the \emph{dynamic} behavior of a system crossing the phase transition. The experiments described here probe the exponents $\nu$ and $z$ only through the combination $\nu/(1+\nu z)$ (see (\ref{eq3}) and (\ref{eq4})). The critical exponent $\nu$ of the (interacting) Bose-Einstein transition has been measured with liquid helium \cite{Pogorelov07} and with ultracold gases \cite{Donner07}, and found to be $\nu\approx 0.67$ \cite{Pelissetto02} (Landau mean-field theory predicts $\nu=1/2$). However the dynamical exponent $z$ in the universality class of the 3D superfluid transition had not been measured before, and critical dynamics provides a way to extract it.

For a given (static) universality class, different dynamical behaviors are possible \cite{Hohenberg77}. The scaling hypothesis, which postulates that the correlation length is the only relevant length scale in the system \cite{Hohenberg77}, allows one to define a universal dynamical behavior, largely independent of the microscopic details of the system. Assuming a generic model for the relaxation of the long-wavelength excitations towards equilibrium, the dynamical critical exponent can be computed. For instance, a Gaussian model in the mean-field approximation leads to $z=2$ for a non-conserved order parameter. In the case of the interacting Bose-Einstein condensation transition, the static universality class is the 3D-XY model. The relevant dynamic universality class is that of the model F \cite{Hohenberg77}, which is also the model for the superfluid transition in liquid helium. It corresponds to a model of an asymmetric planar magnet with a non-conserved order parameter of dimension 2. From the long-wavelength (linear) dispersion relation of the superfluid, one can compute $z=3/2$ \cite{Hohenberg77,Corman16}. 

\begin{figure*}[hbt!!]
    \centering
		\includegraphics[width=2\columnwidth]{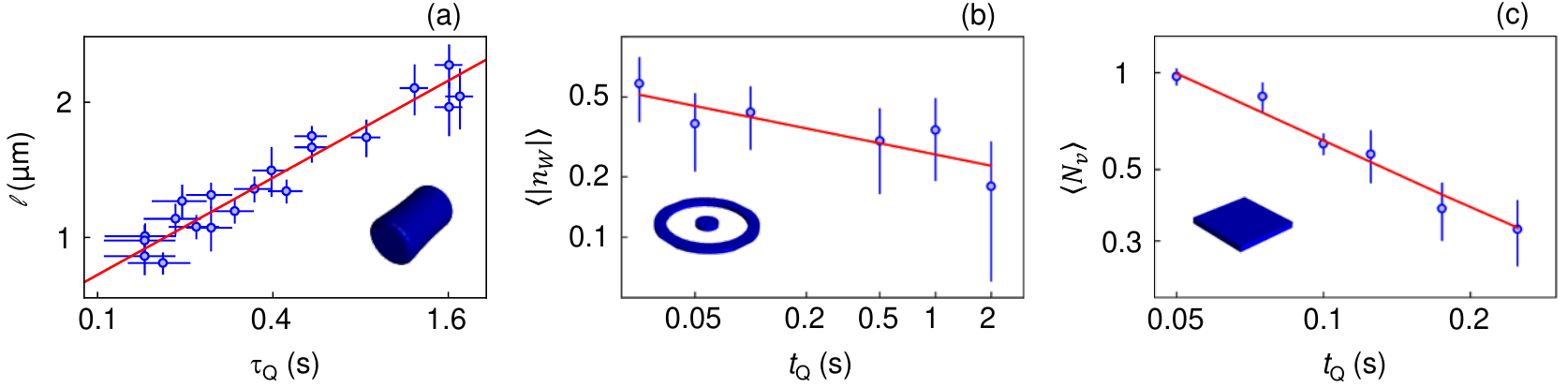}
\caption{
Kibble-Zurek power-law scalings. (a) First-order correlation function measurement of $\ell$ versus quench time $\tau_Q$ for the 3D Bose gas \cite{Navon15}. (b) Mean absolute winding number measured in the ring geometry versus cooling time $t_Q$ \cite{Corman14}. (c) Average number of vortices detected in the 2D bulk configuration as a function of cooling time $t_Q$ \cite{Chomaz15}. This plot does not include the points for slow quenches, for which the number of vortices is approximately constant (see \cite{Chomaz15}). Power-law fits are shown as solid red lines. From \cite{Navon15} \copyright Science AAAS, \cite{Corman14} \copyright American Physical Society and \cite{Chomaz15} \copyright Nature Publishing Group.}
\label{fig4}
\end{figure*}

In the 3D box experiment \cite{Navon15}, the power-law scaling (\ref{eq3}) is obtained from the measurement of $\ell$ for various quench times $\tau_Q$ (see figure \ref{fig4}a). The quantity $\nu/(1+\nu z)$ is measured to be 0.35(4). This value is in good agreement with the F-model value $0.33$ and excludes the mean-field value of $0.25$. Using the established value $\nu=0.67$, one deduces $z=1.4(4)$, which supports the previously unverified expectation that $z=3/2$ for the BEC transition.

%%%%%%%%%% FIGURE EXPOSANTS CRITIQUES %%%%%%%%%%%%
%%%%%%%%%%%%%%%%%%%%%%%%%%%%%%%%%%%%%%%%%%%%%%%%%
%%%%%%%%%%%%%%%%%%%%%%%%%%%%%%%%%%%%%%%%%%%%%%%%%

In the annular geometry of \cite{Corman14}, the average absolute winding number is found to scale as $\langle|n_w|\rangle \propto t_Q^{-\beta}$ with $\beta=0.19(6)$ (see figure \ref{fig4}b). To compare this result to (\ref{eq4}), one has to first determine the appropriate power-law exponent. The typical number of phase domains with independent phases along the annulus is given by $N_\phi=\mathcal{C}/\hat{\xi}$ where $\mathcal{C}$ is the circumference of the annulus. We describe the phase variation along the annulus as a random walk with $N_\phi$ steps. Assuming that a large number of phase domains has been formed we find that the variance of the winding number $n_w$ scales as $\sqrt{N_\phi}$. The average absolute winding number thus varies as $\langle|n_w|\rangle \propto \sqrt{N_\phi}  \propto \hat\xi^{-1/2}$ and we get 
\begin{eqnarray}
\langle|n_w|\rangle  \propto \tau_Q^{-\frac{1}{2}\frac{\nu}{1+\nu z}} . 
\label{eq5}
\end{eqnarray}
The mean-field model (resp. model F) thus predicts $\beta_{\rm MF}=0.125$ (resp. $\beta_{\rm F}=0.167$). Importantly, the result (\ref{eq5}) assumes a large number of phase domains. For very small number of phase domains (and hence for average absolute winding number $\langle |n_w|\rangle \ll 1$) a different power-law scaling is expected, $\langle|n_w|\rangle \propto \hat{\xi}^{-2}$ \cite{Zurek13}. The experiment reported in \cite{Corman14} is not in the limit of large number of phase domains ($\langle |n_w|\rangle \approx 1$) and a small deviation from (\ref{eq5}) is expected \cite{Corman14}. We nevertheless note that the observed behavior and the measured value of power-law exponent are compatible with the KZM prediction. 

Finally, in the bulk quasi-2D geometry \cite{Chomaz15}, it is found that vortices are nucleated according to $\langle N_v \rangle \propto t_Q^{-\gamma}$ with $\gamma=0.69(17)$ (see figure \ref{fig4}c). Theoretically, we expect $\langle N_v \rangle$ to scale as $\mathcal{S}/\hat{\xi}^2$ where $\mathcal{S}$ is the cloud surface. We then obtain
\begin{eqnarray}
\langle N_v \rangle \propto \tau_Q^{-2\frac{\nu}{1+\nu z}} . 
\label{eq6}
\end{eqnarray}
This result has to be compared with the mean-field prediction, $\gamma_{\rm MF}=0.5$ and the F-model one, $\gamma_F=0.67$. The moderate statistics does not allow for a stringent test but the measured power-law behavior is compatible with the KZM prediction, and the exponent, with the F-model prediction. We note that the dynamics of the vortices is neglected in (\ref{eq6}). For low average vortex numbers, the lifetime of the vortices does not modify noticeably the power-law exponent \cite{Donadello16,Chomaz15}, but for larger values of $N_v$, vortex-vortex interactions can lead to a breakdown of the KZ scaling law \cite{Donadello16}. 

For the square and ``target" geometries, the comparison with the predictions for a weakly interacting Bose gas should be considered with care. In a two-dimensional Bose gas, the existence of the BEC transition is prohibited by the Mermin-Wagner theorem. However, the situation described in \cite{Corman14,Chomaz15} does not correspond to a purely 2D regime. The final state of the gas after the quench cooling is a 2D superfluid\footnote{In this context, 2D means that the relevant energy scales of the gas, here the interaction energy $E_{\rm int}$ and the thermal energy $k_B T$, are much smaller than the energy difference $E_\perp$ between the single-particle ground state and the first excited state along the direction perpendicular to the atomic plane.}, but the initial system before the quench is not two-dimensional. The phase transition crossed in this case is not the usual Bose-Einstein condensation nor the Berezinskii-Kosterlitz-Thouless (BKT) transition \cite{Berezinskii71,Kosterlitz73} but a specific transition called transverse condensation \cite{Chomaz15,vanDruten97,Armijo11,RugWay13}. It corresponds to a macroscopic accumulation of the atoms in the ground state of the motion along the transverse direction, driven by Bose stimulation. During transverse condensation the correlation length  quickly increases to a value about the transverse size of the sample \cite{Corman14,Chomaz15} and a freeze-out period is expected around this point in a similar way as for the usual BEC transition. To the best of our knowledge the static and dynamical critical exponents for the transverse condensation transition have not yet been studied.

%%%%%%%%%%%%%%%%%%%%%%%%%%%%%%%%%%%%%%%%%%%%%%%%%%
%%%%%%%%%%%%%%%%%%%%%%%%%%%%%%%%%%%%%%%%%%%%%%%%%%
%%%%%%%%%%%%%%%%%%%%%%%%%%%%%%%%%%%%%%%%%%%%%%%%%%
%%%%%%%%%%%%%%%%%%%%%%%%%%%%%%%%%%%%%%%%%%%%%%%%%%
%%%%%%%%%%%%%%%%%%%%%%%%%%%%%%%%%%%%%%%%%%%%%%%%%%
%%%%%%%%%%%%%%%%%%%%%%%%%%%%%%%%%%%%%%%%%%%%%%%%%%
\section{Prospects}\label{sec8}
We have seen that ultracold atoms are well-suited to study out-of-equilibrium physics and, more particularly, the Kibble-Zurek mechanism. The main advantages with these systems are (i) the possibility to create and manipulate a wide range of non-equilibrium states, (ii) the flexibility in choosing the geometry of the cloud, (iii) the various detection tools that allow the probing of coherence or topological defects. We have also seen that studying these gases can give access to the dynamical critical exponent $z$, which can be difficult to measure in other systems. However these experiments, and previous ones in similar settings, also point out some limitations and challenges to overcome to deepen our understanding of out-of-equilibrium (and even equilibrium) properties of such systems. 

First, whereas these experiments provide a comprehensive picture of several aspects of the KZM, it will be of great interest to  combine all these methods on the same system, namely detecting topological defects in the 3D system or measuring correlations in a 2D system.

Post-quench dynamics should be investigated in more details, especially the formation of topological defects. The detailed mechanism of how they form and evolve starting from a given out-of-equilibrium situation is not well understood and, in several cases, deviations from Kibble-Zurek scaling have been predicted \cite{Biroli10,Su13,McDonald15,Chesler15}. It would be interesting for instance, to implement real-time measurement of the evolution of these defects, as demonstrated in \cite{Serafini15}. Merging of different phase domains during post-quench evolution can also be studied in a controlled way: using custom potentials, it is possible to realize several tens of independent condensates in equilibrium and investigate their merging when removing the barrier separating them, as in the first experiments with few independent clouds \cite{Weiler08}. The Kibble-Zurek mechanism focuses on the power-law scaling because of the associated universality. However, a comprehensive understanding of this out-of-equilibrium situation would require predicting the value of the measured observables (correlation length, density of topological defects) and not only their scaling. 

The possibility to tune the trapping potential landscape should allow the detailed study of the connection between the KZM in homogeneous and inhomogeneous media. This link is controlled by the competition between the velocity of the front with which the critical point propagates in a density-inhomogeneous system as it is quenched, and the speed at which correlations propagate \cite{delCampo14}. By controlling the level of density non-uniformity, one could explore the crossover between these two limiting regimes of the KZM, and the corresponding change of exponents \cite{Pyka13}.

Multi-component uniform quantum gases (with different internal states or even different atomic species) have so far not been produced, and would open up the possibility to explore the homogeneous KZM of more complex symmetry-breaking phase transitions than the U(1)-breaking transition of BEC \cite{Clark16,Sadler06}.

The measurement of power-law behavior usually requires to span the parameters over a large range. In ultracold atom experiments, the typical cooling ramp time is limited, for fast quenches, by the time for the system to reach the temperature imposed from outside (typically 10\,ms) and, for long quenches, by the typical lifetime of the sample on the order of a few seconds. Other cooling methods, such as sympathetic cooling with a large coolant (e.g. another atomic species) may allow one to explore shorter quench times. 

An appealing perspective of these works is also the possibility to study, in strictly 2D systems, the critical dynamics through the BKT transition from a normal state to a superfluid state. It is not a second-order phase transition but it can also be described in an extended framework of the KZM \cite{Jelic11,Dziarmaga14}. More exotic geometries could also be investigated, such as coupled ring condensates, where Josephson vortices are expected \cite{Su13}.

A current limitation to the present experiments is that they are sensitive only to a given combination of $\nu$ and $z$, namely $\nu/(1+\nu z)$. An exciting improvement would be to measure the freeze-out time $\hat{t}$ as a function of quench time $\tau_Q$, which possesses a different scaling:
\begin{eqnarray}
\hat{t} \propto \tau_Q^{\frac{\nu z}{1+\nu z}} . 
\label{eq7}
\end{eqnarray}
This quantity can be deduced from the measurement of $\ell$ along the quench cooling trajectory (see red shaded area in figure \ref{figG1}c) or, as predicted in \cite{Das12} for an annulus geometry, from the delayed increase of the condensed fraction.
Consequently a simultaneous measurement of the power-law exponent for $\hat{t}$ and for $\hat{\xi}$ will yield independent combinations of $\nu$ and $z$ in a single experiment.

\ack We thank J. Dalibard, Z. Hadzibabic, and W.H. Zurek for careful reading of the manuscript and all former and present members of both teams who contributed to the results described here. This work is supported by Region \^Ile-de-France (DIM Nano-K), ANR (ANR-12-BLANAGAFON), ERC (Synergy UQUAM), AFOSR, ARO, DARPA OLE, and EPSRC (EP/K003615/1). N.N. acknowledges support from Trinity College, Cambridge.

\section*{References}
\bibliographystyle{unsrt}
\bibliography{reviewKZbib3}

\end{document}